\documentclass[a4paper]{panl}
\usepackage{cite}
\usepackage{wrapfig}
\usepackage{graphicx}
\usepackage{amssymb}
\usepackage{amsfonts}
\usepackage{amsmath}
\usepackage{longtable}
\usepackage{rotating}
\usepackage{lscape}
\usepackage{epsfig}
\usepackage{multirow}

\usepackage{lineno}

\usepackage{bm}

\originalTeX
\begin{document}

\title{Oscillations of Majorana neutrinos in supernova and CP violation}
\maketitle
\authors{A.\,Popov$^{a,}$\footnote{E-mail: ar.popov@physics.msu.ru},
A.\,Studenikin$^{a,}$\footnote{E-mail: studenik@srd.sinp.msu.ru}}
\setcounter{footnote}{0}
\from{$^{a}$\,Department of Theoretical Physics, \\ Moscow State University, 119991 Moscow, Russia}


\begin{abstract}
Leptonic CP violation is one of the most important topics in neutrino physics. CP violation in the neutrino sector is strongly related to the nature of neutrinos: whether they are Dirac or Majorana particles. In \cite{Popov:2021icg} we have shown that for Majorana neutrinos appearance of nonzero Majorana CP-violating phases combined with strong magnetic field during supernova core-collapse can induce new resonances in neutrino oscillations, for example in $\nu_e \to \bar{\nu}_\tau$ channel. In this paper we further study new resonances in Majorana neutrino oscillations, in particular energy dependence of amplitudes of resonant oscillations. Our findings suggest a potential astrophysical setup for studying the nature of neutrino masses and leptonic CP violation and may be important for future neutrino experiments, such as JUNO, Hyper-Kamiokande and DUNE.
\end{abstract}
\vspace*{6pt}

\noindent
PACS: 13.15.$+$g; 14.60.Pq; 14.60.St

\label{sec:intro}
\section*{Introduction}
Neutrino electromagnetic properties are closely tied with the nature of neutrino mass. In particular, the matrix of neutrino magnetic moments has a different form for the Dirac and Majorana cases \cite{Giunti:2014ixa}, namely

\begin{equation}\label{mag_moments}
	\mu^D = \begin{pmatrix}
		\mu_{11} & \mu_{12} & \mu_{13} \\
		\mu_{12} & \mu_{22} & \mu_{23} \\
		\mu_{13} & \mu_{23} & \mu_{33}
	\end{pmatrix}, \;\;\;
	\mu^M = \begin{pmatrix}
		0 & i\mu_{12} & i\mu_{13} \\
		-i\mu_{12} & 0 & i\mu_{23} \\
		-i\mu_{13} & -i\mu_{23} & 0
	\end{pmatrix},
\end{equation}
where $\mu_{ik}$ are real numbers. Thus, studying evolution of neutrinos in a magnetic field can be a potential way to distinguish between Dirac and Majorana cases. Astrophysical objects possessing strong magnetic fields, such as supernovae, are among most promising setups to probe neutrino electromagnetic properties.

In this paper we study oscillations of Majorana neutrinos in supernova magnetic field and matter. In particular, we focus on CP-violating effects. Previously, CP-violation in oscillations of supernova neutrino was studied in \cite{Balantekin:2007es,Gava:2008rp}. However, in these papers effects of neutrino interaction with a magnetic field were neglected.

\section{Majorana neutrino oscillations in supernova media}
It is well known that neutrino mixing matrix for the case of Majorana neutrinos can be written as follows:
\begin{equation}
	U = U_{PMNS}(\theta_{12}, \theta_{13}, \theta_{23}, \delta) \cdot \text{diag}(e^{i\alpha_1}, e^{i\alpha_2}, 1),
\end{equation}
where $ U_{PMNS}$ is the Pontecorvo–Maki–Nakagawa–Sakata matrix, $\theta_{ik}$ are mixing angles, $\delta$ is Dirac CP-violating phase, and $\alpha_1$ and $\alpha_2$ are Majorana CP-violating phases that only present in case if neutrinos are Majorana particles. In our paper \cite{Popov:2021icg} we have studied the process of Majorana neutrino oscillations in environment peculiar for supernova explosion, in particular strong magnetic field ($10^{12}$ Gauss and stronger) and dense matter. It was shown that presence of nonzero Majorana CP-violating phases can modify patterns of neutrino-antineutrino oscillations induced by the supernova magnetic field.

Majorana neutrino interaction with a magnetic field is described by the following Lagrangian:
\begin{equation}\label{mag_field_int}
	\mathcal{L}_{mag}^M = -\sum_{i,k} \mu_{ik}^M\left[ \overline{(\nu_i^L)^c} \bm{\Sigma}\bm{B} \nu_k^L + \overline{\nu_i^L} \bm{\Sigma}\bm{B} (\nu_k^L)^c \right],
\end{equation}
where $\bm{B}$ is the magnetic field vector and

\begin{equation}
\Sigma_i = \begin{pmatrix}
	\sigma_i & 0 \\
	0 & \sigma_i
\end{pmatrix},
\end{equation}
where $\sigma_i$ are the Pauli matrices.

Majorana neutrino interaction with electrically neutral supernova matter is described by the Lagrangian
\begin{equation}\label{matter_int_majorana}
	\mathcal{L}_{mat}^M = -\sum_{\alpha} V^{(f)}_{\alpha\alpha} \left[\overline{\nu^L_{\alpha}} \gamma_0 \nu_{\alpha}^L - \overline{(\nu^L_{\alpha})^c} \gamma_0 (\nu_{\alpha}^L)^c  \right],
\end{equation}
where 
\begin{equation}
	V^{(f)} = \operatorname{diag} \left(\frac{G_F n_e}{\sqrt{2}} - \frac{G_F n_n}{2\sqrt{2}}, - \frac{G_F n_n}{2\sqrt{2}}, - \frac{G_F n_n}{2\sqrt{2}} \right)
\end{equation}
is the well-known Wolfenstein potential. Here $n_e$ and $n_n$ are electron and neutron number densities of supernova media respectively. Note that due to matter neutrality, proton number density $n_p$ equals to electron number density $n_e$.

To study neutrino oscillations in supernova environment we numerically solve the following equation
\begin{equation}\label{dirac_equation}
	(i\gamma^{\mu} \partial_{\mu} - m_i - V^{(m)}_{ii} \gamma^{0}\gamma_5)\nu_i(x) -\sum_{k \neq i} \big(\mu_{ik}^M \bm{\Sigma}\bm{B} + V^{(m)}_{ik}  \gamma^{0}\gamma_5\big)\nu_k(x) = 0,
\end{equation}
where $V^{(m)} = U^{\dag}V^{(f)}U$ is the matter potential in the massive neutrinos basis, and $n_e$, $n_n$ are electron and neutron number densities respectively. The magnetic moments matrix of Majorana neutrinos $\mu^M$ from Eq. (\ref{dirac_equation}) is antisymmetric and imaginary, and is given by Eq. (\ref{mag_moments}).

The values of neutrino magnetic moments are presently unknown. Currently, the best upper bounds for transition magnetic moments of Majorana neutrinos are given by \cite{deGouvea:2022znk} 
\begin{eqnarray}
	|\mu_{12}^M| &\le& 0.64 \times 10^{-11} \mu_B, \nonumber \\
	|\mu_{13}^M| &\le& 0.75 \times 10^{-11} \mu_B, \;\;\;\; \text{(90\% C.L.)}\\
	|\mu_{23}^M| &\le& 1.1 \times 10^{-11} \mu_B, \nonumber
\end{eqnarray}
where $\mu_B$ is the Bohr magneton.

\begin{figure}[h]
	\begin{minipage}[h]{0.49\linewidth}
		\center{\includegraphics[width=1\linewidth]{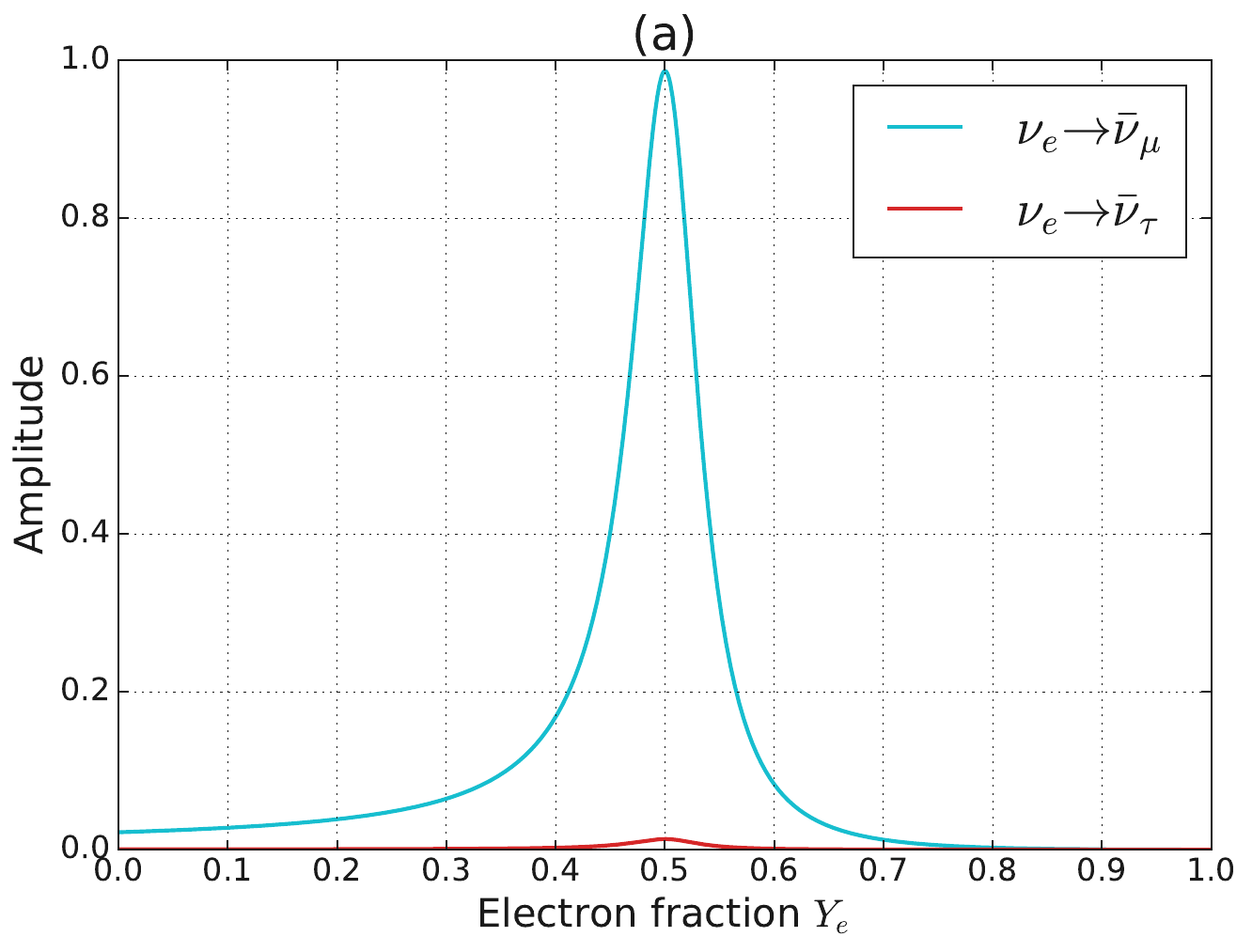}}
	\end{minipage}
	\begin{minipage}[h]{0.49\linewidth}
		\center{\includegraphics[width=1\linewidth]{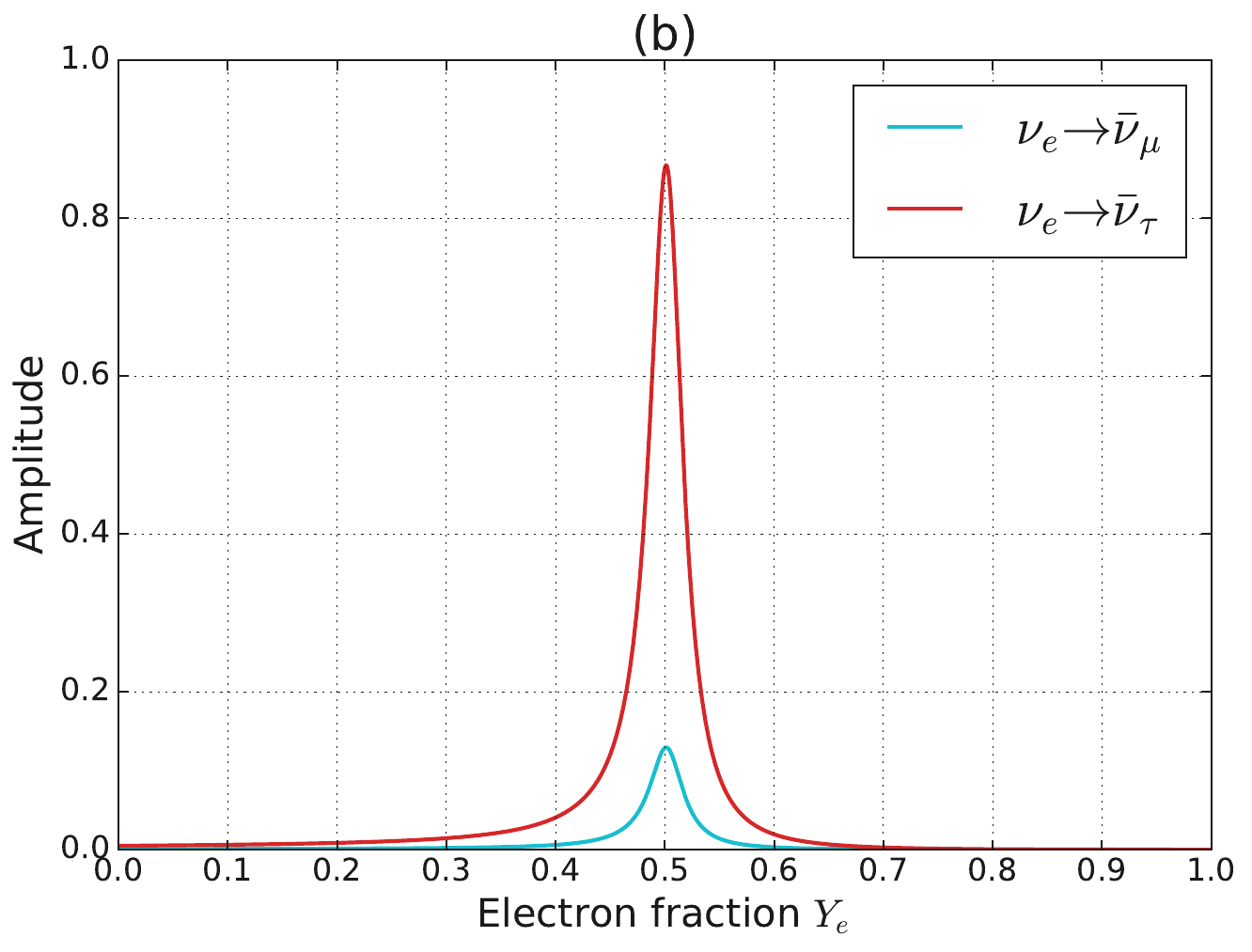}}
	\end{minipage}
	\vspace{-3mm}
	\caption{Amplitudes of Majorana neutrino oscillations in supernova media for $B = 10^{12}$ Gauss as functions of electron fraction $Y_e$. (a) $\alpha_1 = 0$, $\alpha_2 = 0$; (b) $\alpha_1 = \pi$, $\alpha_2 = 0$.}
	\label{fig1}
	\vspace{-5mm}
\end{figure}

Using Eq. (\ref{dirac_equation}), we numerically study the amplitudes of neutrino-antineutrino oscillations. Here we take realistic values for supernova magnetic field $B = 10^{12}$ Gauss, matter density $n_n = 10^{30}$ cm$^{-3}$ and neutrino energy $E_\nu = 12$ MeV. For simplicity we also assume $\mu_{12} = \mu_{13} = \mu_{23} = 10^{-12} \mu_B$, which does not contradict the current experimental results. We study the dependence of the neutrino oscillations probabilities on various parameters, including the electron fraction $Y_e = n_e/(n_n +n_e)$, since it varies significantly along the neutrino path, and the neutrino energy $E_\nu$. Figure \ref{fig1}a shows the maximal probability of $\nu_e \to \bar{\nu}_\mu$ and $\nu_e \to \bar{\nu}_\tau$ oscillations as functions of electron fraction $Y_e$ in supernova media for the case $\alpha_1 = 0$ and $\alpha_2 = 0$. The amplitude of $\nu_e \to \bar{\nu}_\tau$ transition is close to zero, while $\nu_e \to \bar{\nu}_\mu$ oscillations undergo resonant enhancement at $Y_e = 0.5$. This well known phenomenon of resonant spin-flavour conversion was proposed for the first time in \cite{Akhmedov:1988uk,Lim:1987tk}. In Figure \ref{fig1}b we show the amplitudes of neutrino-antineutrino oscillations for the case of nonzero Majorana CP-violating phases, namely $\alpha_1 = \pi$ and $\alpha_2 = 0$. In this case resonant enhancement in $\nu_e \to \bar{\nu}_\tau$ oscillations appears instead of $\nu_e \to \bar{\nu}_\mu$ oscillations at $Y_e = 0.5$. In \cite{Popov:2021icg} we have shown that this new resonance can result in potentially observable phenomena during supernova core-collapse, in particular to modification of $\nu_e$ and $\bar{\nu}_e$ fluxes ratio.

Figure 2 shows amplitudes of resonant conversion $\nu_e \to \bar{\nu}_\mu$ (left) and $\nu_e \to \bar{\nu}_\tau$ (right) at $Y_e = 0.5$ as functions of both Majorana CP-violating phases $\alpha_1$ and $\alpha_2$. We find that $\nu_e \to \bar{\nu}_\mu$ oscillations are present for CP-conserving values, i.e. $\alpha_1,\alpha_2 = 0,2\pi$, while $\nu_e \to \bar{\nu}_\tau$ arise when $\alpha_1$ is close to $\pi$.

\begin{figure}[h]
	\begin{minipage}[h]{0.49\linewidth}
		\center{\includegraphics[width=1\linewidth]{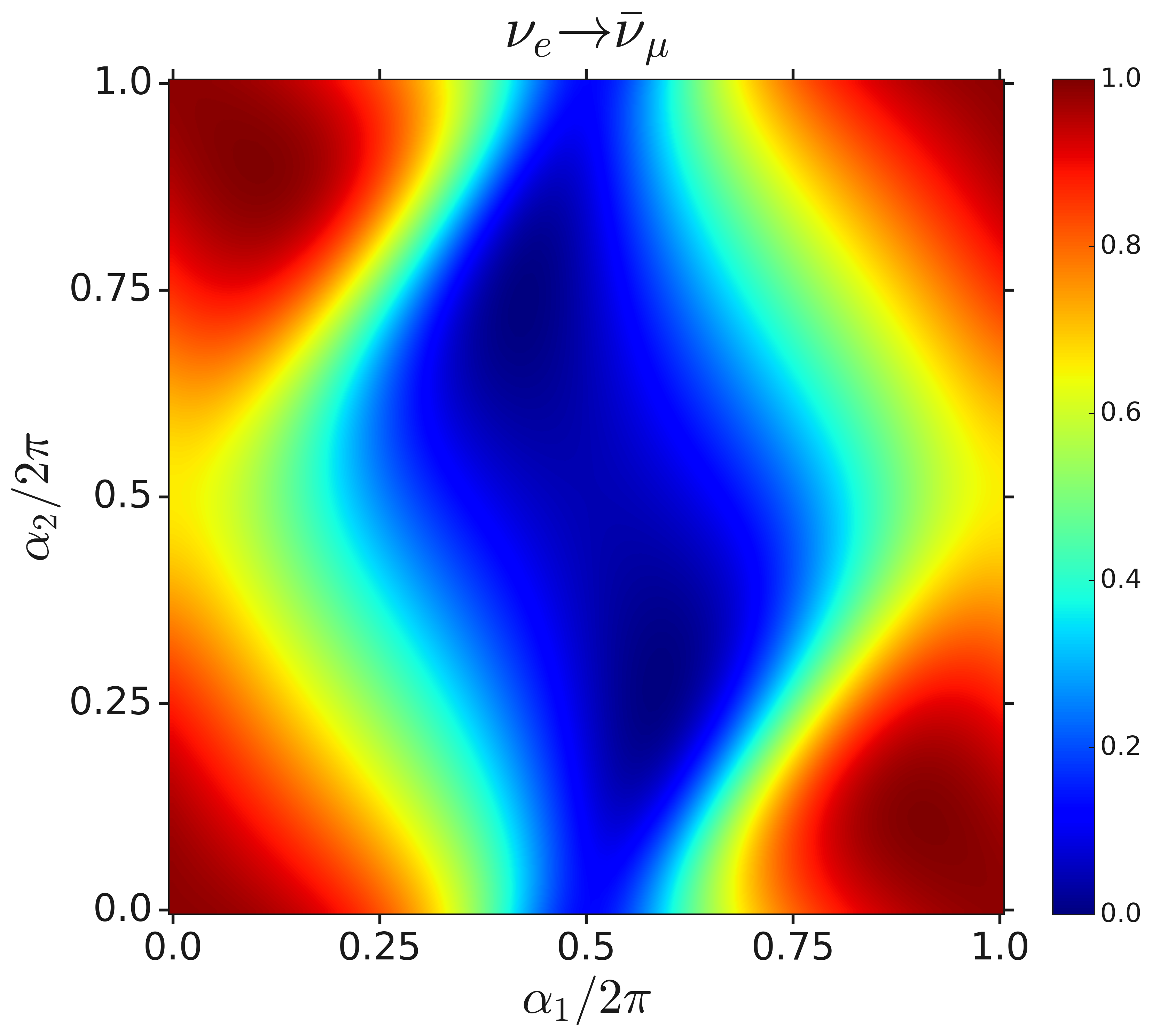}}
	\end{minipage}
	\begin{minipage}[h]{0.49\linewidth}
		\center{\includegraphics[width=1\linewidth]{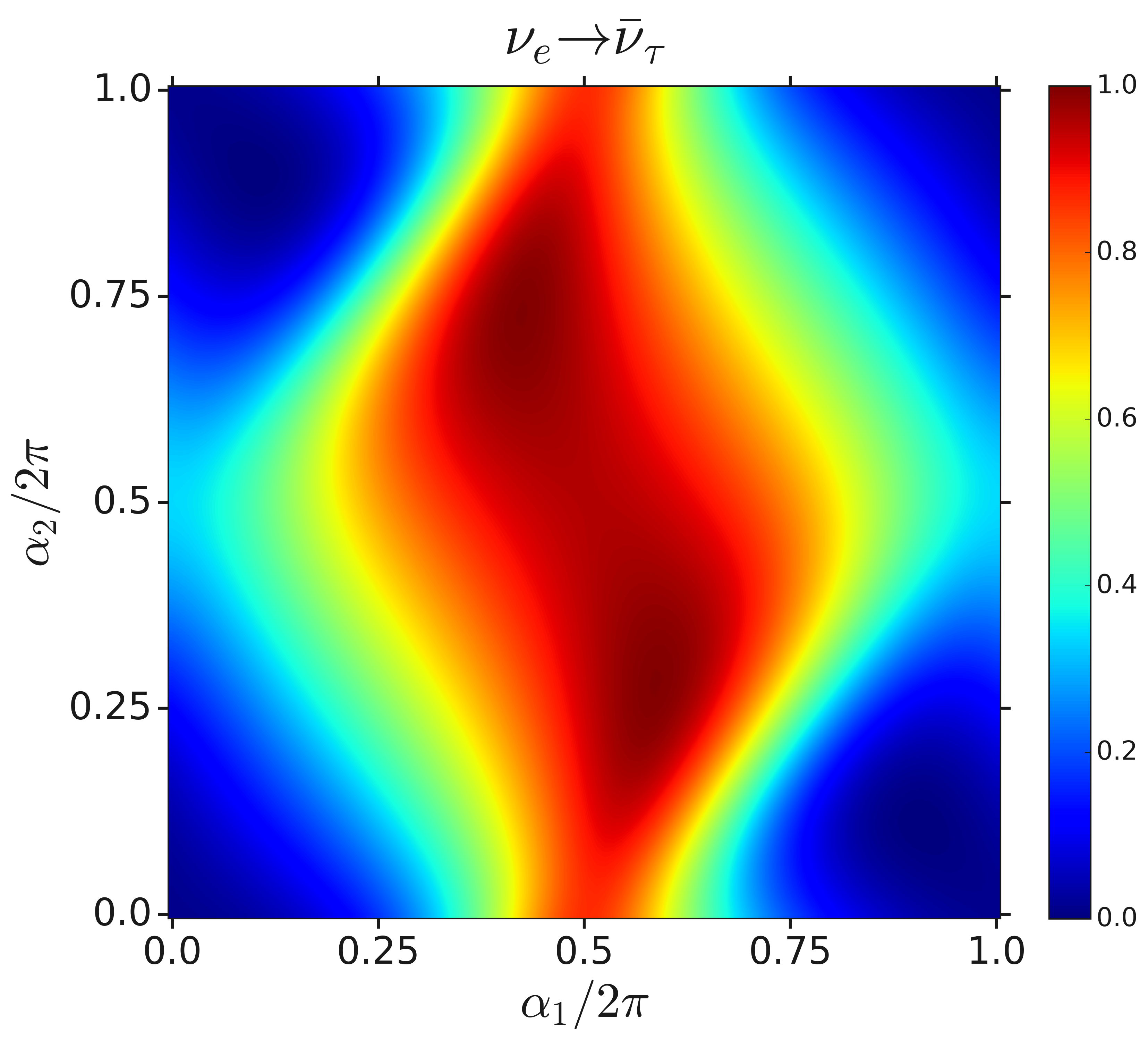}}
	\end{minipage}
	\vspace{-3mm}
	\caption{Amplitudes of resonant conversion $\nu_e \to \bar{\nu}_\mu$ (left) and $\nu_e \to \bar{\nu}_\tau$ (right) at $Y_e = 0.5$ as functions of Majorana CP-violating phases $\alpha_1$ and $\alpha_2$.}
	\label{fig2}
	\vspace{-5mm}
\end{figure}

\begin{figure}[h]
	\begin{minipage}[h]{0.49\linewidth}
		\center{\includegraphics[width=1\linewidth]{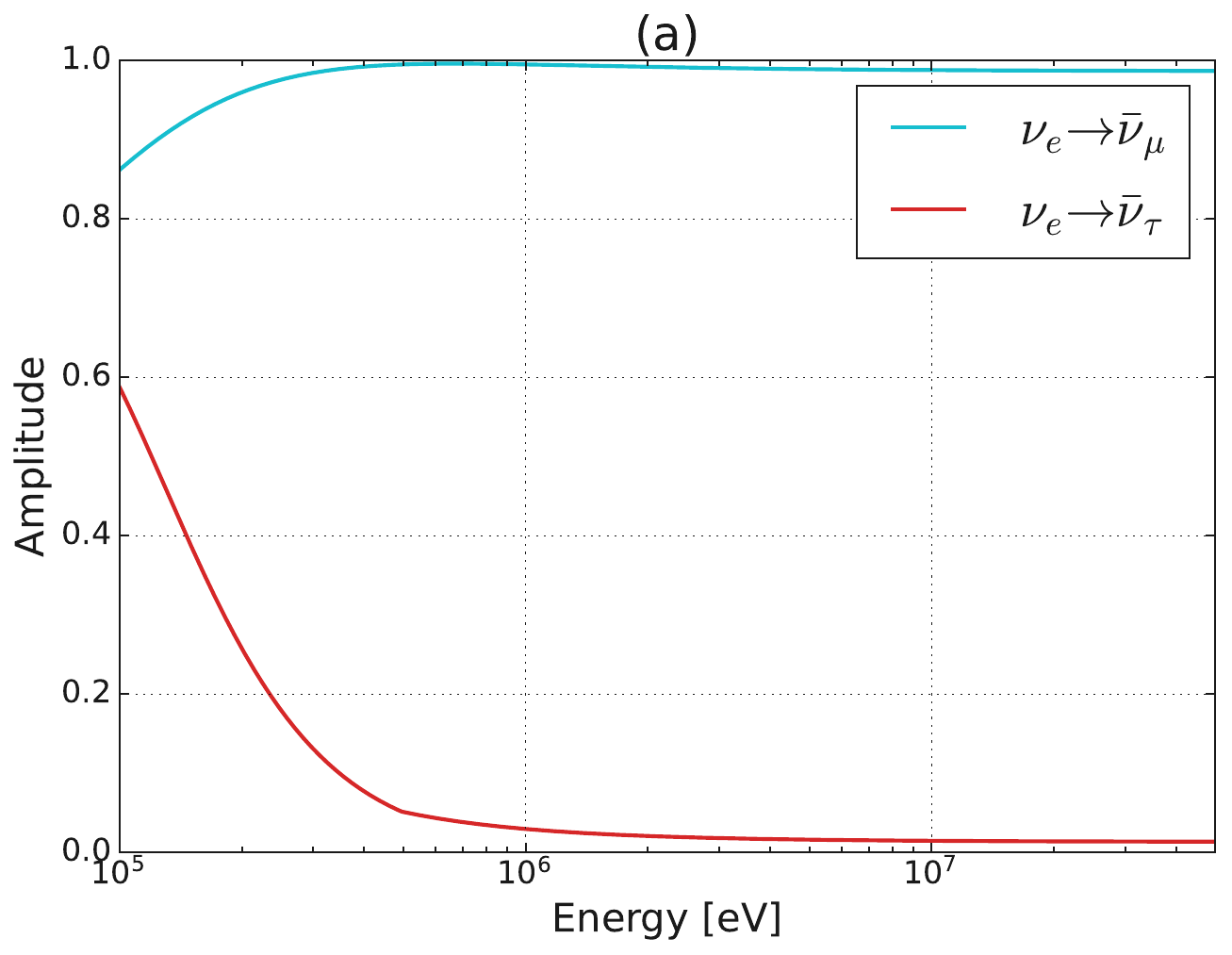}}
	\end{minipage}
	\begin{minipage}[h]{0.49\linewidth}
		\center{\includegraphics[width=1\linewidth]{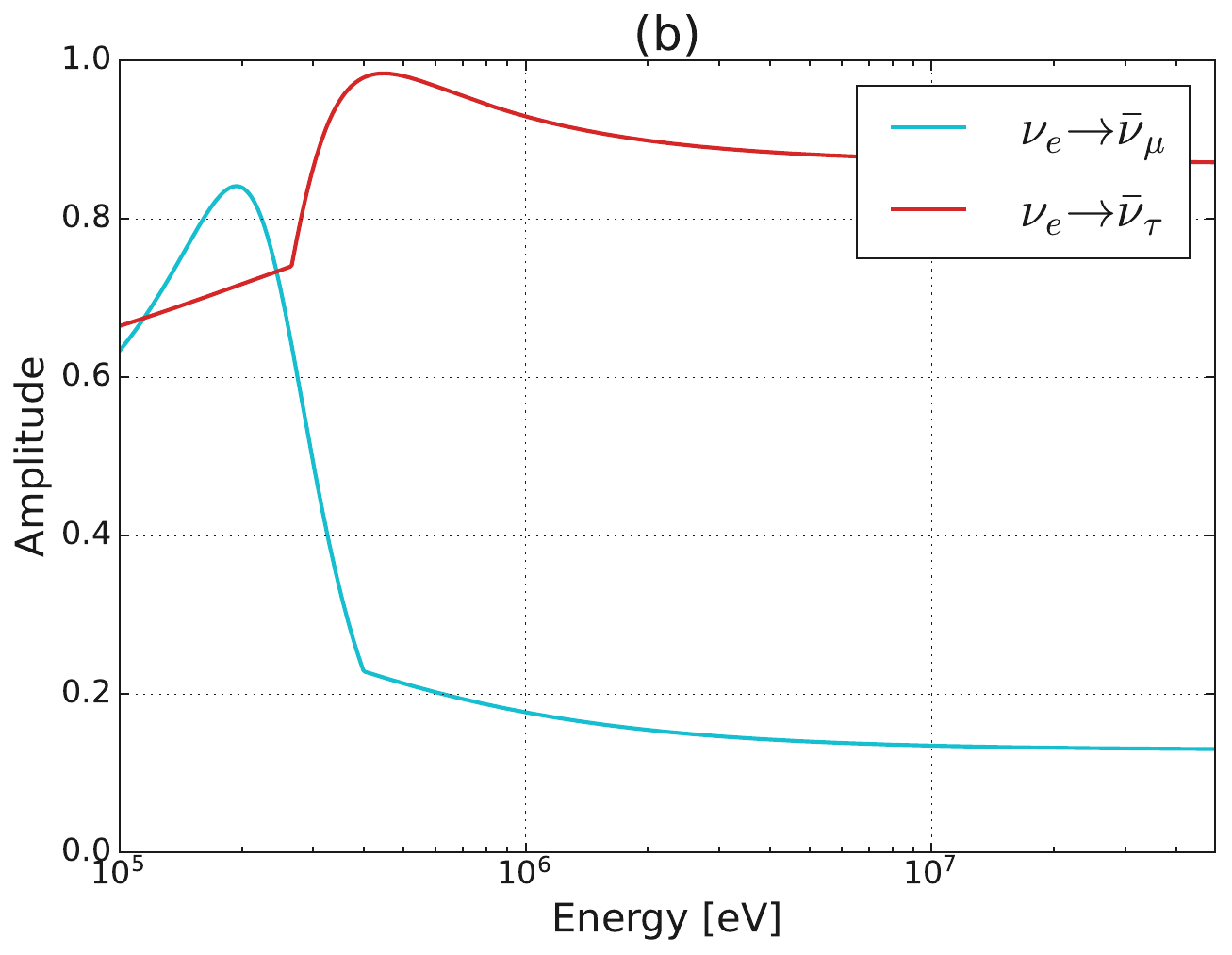}}
	\end{minipage}
	\vspace{-3mm}
	\caption{Amplitudes of resonant conversion $\nu_e \to \bar{\nu}_\tau$ at $Y_e = 0.5$ for $B = 10^{12}$ Gauss as functions of neutrino energy. (a) $\alpha_1 = 0$, $\alpha_2 =0$; (b) $\alpha_1 = \pi$, $\alpha_2 =0$.}
	\label{fig3}
	\vspace{-5mm}
\end{figure}

Throughout \cite{Popov:2021icg} we assumed that neutrino energy is constant $E = 10$ MeV, which is considered to be a typical value for supernova neutrinos. It is also important to study behaviour of new resonance at different neutrino energies. Figure \ref{fig3}a and Figure \ref{fig3}b show the amplitudes of resonant $\nu_e \to \bar{\nu}_\mu$ and $\nu_e \to \bar{\nu}_\tau$ conversions as functions of neutrino energy $E$. We notice that in both cases resonant enhancement is indeed observed for supernova neutrinos with energy of 1 MeV and higher. However, for energies lower that 1 MeV resonant enhancement may disappear.

\section{Conclusion}
In this paper we have studied the process of Majorana neutrino oscillations in supernova media, in particular in a strong magnetic field. We have shown that for certain nonzero values of Majorana CP-violating phases $\alpha_1$ and $\alpha_2$, resonant enhancement of $\nu_e \to \bar{\nu}_\tau$ oscillations can occur in region of supernova media where electron fraction $Y_e = 0.5$. Note that $Y_e = 0.5$ is expected to be realized during supernova core-collapse (see \cite{Buras:2005rp}). We show that this resonant amplification is present for realistic supernova neutrino energies, but it can disappear for energies lower that $\sim$ 100 keV.

The appearance of the considered resonant amplification of neutrino oscillations can lead to a modification of supernova neutrino burst flavour composition. This will lead to a modification of expected event rates from the Galactic supernova explosion in future neutrino experiments, such as JUNO \cite{JUNO_SN}, Hyper-Kamiokande \cite{HK_SN} and DUNE \cite{DUNE_SN}, in particular $N(\bar{\nu}_e)/N(\nu_e)$ ratio. Thus, we expect that supernova neutrino data can be used to probe neutrino magnetic moments, as well as to test the hypothesis of Majorana nature of neutrino mass.

Note that accurate prediction of supernova event rates requires modelling of neutrino fluxes using a specific model of supernova medium, in particular density and magnetic field profiles, and is outside the scope of this paper. In addition, since supernova neutrinos are emitted from the spherical surface called the neutrinosphere, the total neutrino flux is averaged over the emission angle (see for example \cite{Cherry:2013mv}). Averaging over the emission angle will lead to a smearing of the considered effects.

\section*{Acknowledgements}
The work is supported by the Russian Science Foundation under grant No.22-22-00384. The
work of A.P. has been supported by the Foundation for the Advancement of Theoretical Physics and
Mathematics “BASIS” under Grant No. 21-2-2-26-1 and by the National Center for Physics and
Mathematics (Project “Study of coherent elastic neutrino-atom and -nucleus scattering and
neutrino electromagnetic properties using a high-intensity tritium neutrino source”).


\end{document}